\documentstyle[emulateapj,epsfig]{article}
\tighten
\def\spose#1{\hbox to 0pt{#1\hss}}
\def\n{\noindent}

\def\msun{{\rm ~M}_{\odot}}
\def\rsun{{\rm ~R}_{\odot}}
\def\lsun{{\rm ~L}_{\odot}}
\def\mdot{\dot M}
\def\mpy{{\rm ~M}_{\odot} {\rm ~yr}^{-1}}
\def\lta{\mathrel{\spose{\lower 3pt\hbox{$\mathchar"218$}}
     \raise 2.0pt\hbox{$\mathchar"13C$}}}
\def\gta{\mathrel{\spose{\lower 3pt\hbox{$\mathchar"218$}}
     \raise 2.0pt\hbox{$\mathchar"13E$}}}

\def\be{\begin{equation}}
\def\ee{\end{equation}}

\begin{document}

\title {The Structure and Evolution of Circumbinary Disks in Cataclysmic
Variable Systems}
\author{Guillaume Dubus\altaffilmark{1},  Ronald E. Taam\altaffilmark{2},
and H. C. Spruit\altaffilmark{3}}

\altaffilmark{1}{California Institute of Technology, MC 130-33, Pasadena,
CA 91125}

\altaffilmark{2}{Department of Physics \& Astronomy, Northwestern
University, Evanston, IL 60208}

\altaffilmark{3}{Max-Planck-Institut f\"ur Astrophysik, Postfach 1317,
D-85741 Garching, Germany}

\begin{abstract}
We investigate the structure and evolution of a geometrically thin
viscous Keplerian circumbinary (CB) disk, using detailed models of
their radiative/convective vertical structure.  We use a simplified
description for the evolution of the cataclysmic binary and focus on
cases where the circumbinary disk causes accelerated mass transfer
($\gta 10^{-8} \mpy$).  The inner edge of the disk is assumed to be
determined by the tidal truncation radius and the mass input rate into
the disk is assumed to be a small fraction ($10^{-5}$--$0.01$) of the
mass transfer rate.  Under the action of the viscous stresses in the
disk the matter drifts outward with the optically thick region
extending to several AU.  The inner part of the disk is cool with
maximum effective temperatures $\lta 3,000$ K while the outermost
parts of the disk are $\lta 30$ K and optically thin.  We calculate
the effects of thermal instability on a sufficiently massive CB disk.
It leads to outbursts reminiscent of those in thermally unstable
accretion disks, with the instability remaining confined to the inner
regions of the CB disk.  However, for most of the evolutionary
sequences the surface densities required to trigger instability are
not reached.  The spectral energy distributions from circumbinary
disks are calculated, and the prospects for the detection of such
disks in the infrared and submm wavelength regions are discussed.
\end{abstract}


\section{INTRODUCTION}

Over the last decade evidence has been accumulating suggesting the
presence of gaseous matter beyond the orbit of the components in
cataclysmic variable binary systems (CVs). Its existence has been
inferred from P-Cygni profiles in nova-like variables, super soft
sources, and dwarf novae in outburst (see, for example, Deufel et
al. 1999).  In addition, the observation of narrow widths of
single-peaked emission lines in SW Sex stars (see Thorstensen et
al. 1991; Hellier 2000) and low velocity spectral features in AM CVn
which do not follow the orbital motion of the binary system (see
Solheim \& Sion 1994) provide further support for this interpretation.
Such a gaseous component may take the form of an extended shell or a
flattened distribution of matter surrounding the binary system.
Assuming that some fraction of this matter remains bound to the system
and condenses into the orbital plane, in analogy to the compression of
a radiation-driven wind into the equatorial plane in Be-stars
(Bjorkman \& Cassinelli 1993, Bjorkman \& Wood 1995), a circumbinary
disk may form.  The presence of such a CB disk, as proposed by Spruit
\& Taam (2001), can provide an attractive explanation for some of the
discrepancies between observations and the current theory of evolution
of CVs.

The importance of a CB disk for the evolution of a CV lies in the
angular momentum it can extract from the binary, leading to mass
transfer rates exceeding those due to magnetic braking and/or
gravitational radiation.  The binary exerts gravitational torques on
the external disk, thereby transferring some of its orbital angular
momentum.  If some fraction of the mass lost from the main
sequence-like secondary of the system forms a CB disk, a feedback
process can operate to accelerate the binary evolution. As shown in
the detailed computations by Taam \& Spruit (2001) the evolution of
the system can be significantly accelerated and the mass transfer rate
from the secondary towards the white dwarf, $\dot{M}_{\rm 2}$,
elevated to rates as high as $10^{-7} \mpy$.

In previous studies, we have focused on the response of the mass
losing component and on the evolution of the binary system resulting
from the influence of the CB disk, using a one zone approximation for
its vertical structure.  Here, we report on the results of detailed
numerical calculations for the vertical structure and radial evolution
of such disks surrounding a model CV.  In order to determine the
observability of such disks, their spectral energy distributions are
also presented.  The assumptions of the model computations are
described in the next section. In \S3 we describe the evolution of a
CB disk neglecting angular momentum losses from the binary
system. Results with the binary evolution taken into account are shown
in \S4 and several spectral energy distributions for some illustrative
systems are presented in \S5. We summarize and conclude in the last
section.

\section{MODEL}

We assume that the CB disk is axisymmetric and geometrically thin such
that matter rotates at the Keplerian speed. The formulation of the
equations for mass, angular momentum, and energy transport is as
described in Hameury et al. (1998).  Models for the
radiative/convective vertical structure have been calculated and
tabulated in a grid for a range of radii, mid-plane temperatures, and
column densities. The global structure and evolution of the disk
follows from the coupling of this grid to the time dependent transport
of mass, energy, and angular momentum in the radial direction.  The
viscous stress tensor, $\tau_{r \phi}$, is assumed to be equal to
$\alpha P$ where $\alpha$ is the viscosity parameter introduced by
Shakura \& Sunyaev (1973) and $P$ is the local pressure in the disk.
The calculations of the disk differ from those of Taam \& Spruit
(2001) where the vertical structure of the disk was calculated in a
one zone approximation.  Furthermore, we take into account departures
from local thermal equilibrium and include convective energy
transport.  The latter is treated in the mixing length approximation
with the ratio of mixing length to pressure scale height taken to be
equal to 1.5, where the functional form for the pressure scale height
is chosen to ensure that it does not exceed the vertical scale height.

Optically thin regions above the disk photosphere are included in the
computation and are treated in the grey approximation. The input
physics includes an equation of state of matter in which the
ionization/dissociation balance of hydrogen is taken into account.
The opacities are taken from Bell et al. (1997) and include
contributions from iron poor opacities at temperatures ($T \lta 1500$
K) where dust particles exist (Henning \& Stognienko 1996). It is
assumed that the composition of the material corresponds to a solar
mix ( $X = 0.7$, $Y = 0.28$, and $Z = 0.02$).

The boundary conditions are formulated such that both the inner radius
and outer radius of the disk can vary in time.  At the inner radius,
it is assumed that the mass input rate into the CB disk is a constant
fraction of the mass transfer rate from the secondary $\dot{M}_{\rm
i}=-\delta \dot{M}_{\rm 2}$.  Although the mass input into the disk
may extend over a range of radii, we have adopted the simplest
approach and assumed that the mass is deposited at its innermost edge.
The condition that this input mass flux is carried outward by viscous
diffusion serves as an inner boundary condition, which is formulated
as a relation between $\Sigma$ and its gradient. The specific angular
momentum of the mass fed into the disk is taken to be equal to that of
a Kepler orbit at the inner edge.  The mass deposited is thus neutral
in an angular momentum sense.  In addition, it is assumed that the
radial temperature gradient vanishes.  Our treatment differs from that
adopted in Pringle (1991) in that we have a mass input into the disk
which is absent in the formulation developed by Pringle (1991). In the
latter equivalent approach, the radial velocity at the inner edge is
assumed to vanish.  The column density and the mass outflow rate at
the outer boundary are set to $10^{-2}$~g~cm$^{-2}$ and zero
respectively.  With these treatments of the boundaries, the disk
structure and evolutionary equations are solved in a fully implicit
fashion on an adaptive grid (typically with 400 zones).  For a full
description of the numerical method, see Hameury et al. (1998).

Since the primary focus of this investigation is on the structure of
the CB disk, the evolution of the binary system is treated in a
simplified approximation wherein the response of the mass losing star
is modeled assuming homology (Spruit \& Ritter 1983; see also Spruit
\& Taam 2001). The mass transfer process is assumed to be
conservative.  That is, a very small fraction, $\delta < 0.01$, of the
mass lost from the donor star, which itself is assumed to be small
compared to the mass transferred to its companion, is assumed to be
deposited into the CB disk.

Angular momentum loss associated with magnetic braking is
assumed to take place on a fixed timescale $t_{\rm w}$. The
contribution of the CB disk to the angular momentum loss from the
binary is given by the viscous torque at the inner edge of the disk,
\be \dot{J}_{\rm d}= 3\pi (R_{\rm i}/a)^{1/2}\Omega_0 a^2 \nu_{\rm
i}\Sigma_{\rm i}, \ee where $\Omega_0$ is the orbital frequency of the
binary system, $a$ is the orbital separation of the binary components,
and $R_i$, $\nu_i$, and $\Sigma_i$ are the radius, viscosity, and
column density at the inner edge of the CB disk respectively. The
evolution of the system follows from the total angular momentum loss,
(see Spruit \& Taam 2001):
\be
-\dot{J}/J=t^{-1}_{\rm w}+3\pi(R_{\rm i}/a)^{1/2}(1+q)/M_2\,\nu_i \Sigma_i
\ee
It is assumed that the gravitational torque is concentrated near the
inner edge of the disk and that the radius of the inner edge is
approximately equal to 1.7 times the orbital separation.  This latter
assumption is based on the work of Artymowicz \& Lubow (1994) who show
that the ratio of the inner radius of the CB disk to the orbital
separation is relatively insensitive to the mass ratio of the binary
system, q, due to the steep decrease of the tidal torques with
distance.

\section{CB DISKS IN NON-EVOLVING SYSTEMS}

As a first step we computed several CB disk models in which we
neglected all the angular momentum losses from the binary system. The
binary evolution is thereby frozen with the inner radius $R_i$ and
mass input rate $\dot{M}_{\rm i}$ (instead of $\delta$) set as
constant parameters. Our calculations start with a very low density
($\Sigma=0.01$~g~cm$^{-2}$) ring of matter at the inner edge, and we
follow the expansion of the disk as matter gradually piles up and
diffuses outwards.

Fig.~1 is an example of the evolution of the column density at the
inner radius, $\Sigma_{\rm i}$, and of the outer radius, $R_{\rm d}$,
for a CB disk with $\alpha=0.001$, $R_{\rm i}=10^{11}$~cm,
$\dot{M}_{\rm i}=-10^{14}$~g~s$^{-1}$
(the negative value indicating
mass is flowing away from the binary) and a binary mass $M$ of 1
$\msun$.  The outer disk radius as shown in Fig.~1 is defined as the
radius at which the disk becomes optically thin ($\tau=1$). After a
transitory period lasting at most a few hundred years (viscous
timescale of the inner edge), both $\Sigma_{\rm i}$ and $R_{\rm d}$
increase gradually in a roughly self-similar, power law-like
fashion. This behaviour is independent of the initial setup and, to a
large extent, independent of the outer boundary conditions on
$\dot{M}$ and $\Sigma$. Increasing the mass input rate speeds up the
evolution with higher values of $\Sigma_{\rm i}$ reached sooner.
Changing $\alpha$ has a weaker influence with lower viscosities
leading to slightly higher values of $\Sigma_{\rm i}$. Both of these
effects can be seen in Figs.~3-6 which take the binary evolution into
account and are discussed in the next section.

The evolution of the column density, mid-plane temperature, effective
temperature, ratio of scale height to radius, and mass flow rate are
illustrated as a function of radius and time in Fig.~2.  It is evident
that the column densities continually increase at a given radius and
reach densities as high as $10^4$~g~cm$^{-2}$ at the inner disk
edge. The point at which the optical depth equals unity (denoted as a
solid dot), corresponding to $R_{\rm d}$ in Fig.~1, also increases
with time and the optically thick disk reaches a size of about 10 AU
at the end of the calculation.  The mid-plane temperatures (solid
line) and the effective temperatures (dashed line) both increase with
time, reflecting the increasing surface density and energy dissipation
in the disk associated with the increasing mass flow rate.  The
mid-plane (effective) temperatures decrease monotonically with
distance, from 10000 (3000)~K in the innermost regions to 20 (10) K
where the disk becomes optically thin. However, we note that such
densities and temperatures are reached only after a Hubble time, at
which point the neglect of angular momentum losses from the binary is
not realistic. The disk remains geometrically thin throughout the
evolution, with the ratio of pressure scale height to radius in the
optically thick region, $H/R$, $< 0.05$.  We find that $H/R$ does not
increase uniformly with radius, so that the outer disk region is in
the shadow of the inner region, reminiscent of the protostellar
circumstellar disks studied by Bell et al. (1997).

The simple self-similar CB disk model of Spruit \& Taam (2001), which
assumes that the viscosity is a linear function of $R$ only,
$\nu=\nu_{\rm i}(R/R_{\rm i})$, gives for the inner regions of the
disk: \be \Sigma(R,t)\approx\left({t \over {t_{\rm vi}}} \right)^{1/2}
{(-\dot{M}_{\rm i}) \over {3\pi \nu_{\rm i}}} \left( {R\over R_{\rm i}}
\right)^{-3/2} \ee where $t_{\rm vi}=R_{\rm i}^2/\nu_{\rm i}$ is the
viscous timescale at the inner edge of the disk. This implies a
scaling for $\Sigma_{\rm i}\propto \dot{M}_{\rm i}/\sqrt{\alpha}$,
which we find to be roughly reproduced by the numerical results (see
e.g. the early evolution of the disks in Figs.~5-6).

\section{CB DISKS IN EVOLVING SYSTEMS}

Evolutionary sequences were calculated for model CV systems consisting
of a secondary component of 0.55 $\msun$ and a primary white dwarf
component of 0.95 $\msun$.  The disk calculations were performed for
several fractional mass input rates into the CB disk $\delta \lta
0.01$, viscosity parameters $\alpha \lta 0.01$, and time scales for
angular momentum loss by magnetic braking $t_{\rm w} \lta 10^{10}$
yrs.

\subsection{Acceleration of the binary evolution}

The results of an evolutionary sequence with parameter values $\delta
= 0.005$, $t_{\rm w} = 2\times 10^8$ years, and $\alpha = 0.001$ are
illustrated in Figs.~3 and 4. The CB disk initially evolves in the
same fashion as described in the previous section. Angular momentum
losses by magnetic braking then start changing the system parameters
and the system evolves in a standard way, with $-\dot{M}_{\rm i}$
decreasing together with the mass of the secondary. The CB torque
begins to dominate the evolution at $t\approx 20$~Myr, when the mass
input rate into the CB disk is approximately at its minimum,
$\dot{M}_{\rm i}\approx -3 \times 10^{-11} \mpy$ ($\dot{M}_{\rm
2}\approx6\times 10^{-9} \mpy$). The mass input rate is now
proportional to the surface density since
$\dot{M}_{\rm i}/M_2\sim \dot{J}_{\rm i}/J\sim -(\nu \Sigma)_{\rm
i}/M_2$, leading to a runaway on a timescale comparable to that on
which the CB torque becomes dominant (Spruit \& Taam 2001). The
acceleration is manifest
in Fig.~3 as an abrupt increase of $-\dot{M}_{\rm i}$ to $\gta 10^{-9}
\mpy$, leading to an increase in the orbital period as a result of the
tendency for the secondary to expand as it departs significantly from
thermal equilibrium.

By the end of the calculation at $t \approx 65$~Myr the orbital
period is $\sim$ 20~hr and the secondary has $M_2\approx 0.03
\msun$. The response of the star to mass loss certainly deviates from
homology during the runaway but there is little doubt that it will end
with the dissolution of the secondary. Qualitatively similar results
were obtained by Taam \& Spruit (2001) .

\subsection{Influence of the different parameters}

The influence of changing $\alpha$ is also shown in Fig.~3. The CB
torque becomes dominant when $\dot{J}_{\rm d}\gta\dot{J}_{\rm w}$.
Combining Eq.~1-3, we find this happens when (Spruit \& Taam 2001) \be
\dot{J}_{\rm d}/J\sim(t/t_{\rm vi})^{1/2} \dot{M}_{\rm i}/M_2\sim
-t^{-1}_{\rm w}\ee For a given magnetic braking torque and
$\dot{M}_{\rm i}$, the disk torque dominates at $t\propto t_{\rm vi}
\propto 1/\alpha$. This scaling is consistent with the numerical
results of Fig.~3.  Note that for very low values of $\alpha$ ($\lta
0.0001$ for $\delta=0.005$ and $t_{\rm w}=0.2$ Gyr) the CB torque
would be entirely negligible during the course of the evolution.  It
should be pointed out that we have implicitly assumed that $\alpha$ is
constant throughout the disk.  It is possible that $\alpha$ may
decrease significantly at large radii (see Gammie \& Menou 1998) where
the magnetic Reynolds number is decreased below values necessary for a
magnetic dynamo to operate. In this case, the size and the properties
of the outer disk would be modified from that calculated here.
However, the extent of the magnetically decoupled region is model
dependent as it is very sensitive to the ionization fraction in the
disk (see Fromang, Terquem, \& Balbus 2002).

Changing the magnetic braking timescale $t_{\rm w}$ changes
$\dot{M}_{\rm i}$ since the initial rate is set by $\dot{M}_2/M_2\sim
\dot{J}/J\approx t^{-1}_{\rm w}$ (Eq.~2). Using this in Eq.~4, we find
the CB torque should thus start dominating the evolution at roughly
the same time regardless of $t_{\rm w}$. This is confirmed by the
calculations shown in Fig.~5 in which the disk torque dominates after
$t\approx 20$~Myrs for $t_{\rm w}=$ 0.2, 1 and 5 Gyrs. Note that
$\Sigma_{\rm i}$ at $t\approx 20$~Myrs decreases with increasing
$t_{\rm w}$ since $\Sigma_{\rm i}\propto -\dot{M}_{\rm i} \propto
t_{\rm w}^{-1}$ during the early magnetic braking-dominated evolution.
The conditions for the
runaway should be independent of $t_{\rm w}$ and we indeed find the
disks in Fig.~5 to have similar properties when it occurs
($\Sigma_{\rm i}\approx 40000$~g~cm$^{-2}$, $R_{\rm d}\approx 10$~AU).

Fig.~6 illustrates the influence of changing the fraction of the mass
transfered to the CB disk. For $\delta\gta0.01$ the system dissolves
in $\lta 1$~Myr suggesting that systems with high values of $\delta$
are unlikely to be observed. The timescale on which the disk torque
dominates dramatically increases as $\delta$ becomes smaller. The
effect is much stronger than decreasing $\alpha$ as already noted by
Taam \& Spruit (2001). For values of $\delta \lta 0.0001$ the disk
torque becomes important on timescales longer or comparable to the
magnetic braking timescale $t_{\rm w}$. The secondary has then already
lost most of its mass so that the CB disk has no impact on the binary
evolution. Being smaller and less massive, the CB disk is also fainter
and more difficult to detect (see \S5).

Taam \& Spruit (2001), using a detailed evolutionary code for the
secondary but a simpler disk model, found that the CB torque was
negligible when $\delta \lta 0.01$.  The present calculations indicate
that disks formed with $\delta \gta 0.0001$ can still influence the
binary.  The discrepancy can be traced to the assumptions made in the
different models.  In Taam \& Spruit (2001), the column densities at
the inner edge of the disk are very close to the lower turning point
of the S curve ($\Sigma_{\rm max}\approx 4000$~g~cm$^{-2}$ at
$R=10^{11}$~cm with $M=1 \msun$ and $\alpha=0.001$, see Fig.~1 of Taam
\& Spruit). The specific interpolation scheme between the two stable
branches which they used forces the column density to stagnate around
$\Sigma_{\rm max}$ once this value is reached. The detailed vertical
structure computations, which include convection, show that the actual
value of $\Sigma_{\rm max}$ is an order of magnitude higher than in
the one zone radiative model ($\Sigma_{\rm max}\approx
40000$~g~cm$^{-2}$ with the same parameters as above using Eq. 32 of
Hameury et al. 1998 ; see Fig.~2 of Lasota 2001 for a discussion of
the importance of convection on $\Sigma_{\rm max}$). Much higher
densities can be reached on the cold branch and hence higher torques
can be achieved during the evolution for the same $\delta$ and
$\alpha$. On the other hand, the approximate treatment for the
response of the donor to mass loss for the binary evolution in the
present calculation may artificially decrease the minimum $\delta$
needed to have a dominant CB torque.

\subsection{Can CB disks become thermally unstable ?}

Depending upon the evolution of the CB disk, $\Sigma_{\rm i}$ may
become higher than the critical column density, $\Sigma_{\rm max}$,
above which a cold, optically thick and geometrically thin disk
becomes thermally and viscously unstable.  The instability is due to
the large change in opacity which accompanies hydrogen ionization (see
Lasota 2001 for a review). Such a CB disk would present limit cycle
oscillations analogous to those which occur in the accretion disks of
dwarf novae.

While the linear stability problem is the same for the accretion disk
and the CB disk, one should expect some differences to show up in the
nonlinear development.  In the accreting case, the increased mass flux
in the hot state propagates inward into a decreasing volume,
amplifying its effects. The instability in a CB disk starts at the
inner edge and propagates outward into an increasing volume. One might
thus expect its effects to be milder, and the transition front to
stall at some moderate distance from the inner edge. To check on these
expectations we have followed the instability in a few cases. But
first, we investigate for which parameters the evolving binary with a
CB disk can reach instability.

We combine the expression for $\Sigma(R_{\rm i},t)$ (Eq.~3) with that
for $\Sigma_{\rm max}$ (Eq. 32 of Hameury et al. 1998) to estimate the
time needed to reach this critical density, assuming the binary
evolution is frozen (constant $R_{\rm i}$ and $\dot{M}_{\rm i}$). The
central temperature at the critical density is $T_{\rm c}\approx
20000$~K (Hameury et al. 1998). We use this temperature to get an
estimate of $\nu_{\rm i}\approx\alpha c^2_{\rm s} / \Omega_{\rm K}$,
where $c^2_{\rm s}$ and $\Omega_{\rm K}$ are respectively the sound
speed and Keplerian frequency at the inner edge. We find \be t\approx
6\times 10^{10}~~\alpha_{0.001}^{-0.7}~M_1^{-1.2}~\dot{M}_{\rm i,
14}^{-2}~R_{\rm i, 11}^{5.7}{\rm ~~~~years}\ee where $\alpha$ is in
units of 0.001, $\dot{M}_{\rm i}$ is in units of $-10^{14}$~g~s$^{-1}$,
$M$ is in solar masses and $R_{\rm i}$ is in units of $10^{11}$~cm.
The results from a few numerical tests show a shallower dependence on
$\dot{M}_{\rm i}$ (-1 instead of -2) but this is still a fair
order-of-magnitude approximation.  The steep dependence on $R_{\rm i}$
implies that instability can be reached within a Hubble time only if
the binary is sufficiently tight during most of its evolution.

In the calculations of Figs.~3-6, the CB disk never reaches the
critical densities required to become unstable.  For reasonable
assumptions on $M_1$ (0.6-1.0 $\msun$) and $M_2$ (0.1-0.6 $\msun$) the
CB disk has an initial $R_{\rm i}$ of $\approx 0.01$~AU for
which the critical density $\Sigma_{\rm max}$ is of the order of
$10^5$~g~cm$^{-2}$ for $\alpha=0.001$. The density in the disk does
not build up fast enough to trigger the instability before the binary
starts evolving away from the initial conditions.  Unrealistically
high values of $\delta\gta 0.1$ would be required to quicken the
density buildup. At these rates, the binary would also dissolve in a
very short time.

If magnetic braking dominates during the evolution, the CB disk is
necessarily not very massive and $\Sigma_{\rm i}$ stays far from its
critical value.  If the disk torque dominates, the mass transfer rate
and $\Sigma_{\rm i}$ increase quickly, particularly during the
runaway.  However, under the present assumptions, the secondary
expands as it rapidly sheds mass onto the white dwarf (Eq.~14 of
Spruit \& Ritter 1983).  The runaway is thus accompanied by an
increase of the orbital period, and hence of $R_{\rm i}$ (see Fig.~3),
which prevents $\Sigma_{\rm i}$ from becoming greater than
$\Sigma_{\rm max}$ ($\propto R_{\rm i}$).

The adiabatic mass-radius exponent is -1/3 for a sequence of
homologous stars; in this respect they behave like convective stars
(Spruit \& Ritter 1983). This limits the freedom of the model. We have
relaxed this in a test calculation in which the mass-radius
coefficient of -1/3 as it appears in the model was increased to
+1/2. The other parameters are $\alpha=0.005$ and $\delta=0.005$
(Fig.~7). The global evolution is very similar to the one shown in
Fig.~4 except that the secondary contracts in response to the runaway
mass transfer at $t\approx 30$~Myr. The orbital period decreases to
$\sim 2$~hours, bringing the inner edge of the CB disk to 0.007 AU
($M_2\approx 0.085 \msun$, $R_2\approx 0.16 \rsun$). The inner column
density $\Sigma_{\rm i}$ is then greater than the critical density and
the CB disk, which is now $\sim 0.003 \msun$, becomes unstable.

We assume that, when the CB disk is hot, the viscosity coefficient
$\alpha_{\rm hot}$ increases to 0.01 by analogy with the models of
dwarf novae outbursts. The inner region of the CB disk cycles between
a hot and a cold state in a complex fashion on a timescale of $\sim
15$~years, with the hot state lasting about a year (see Fig.~7).
Changes in the CB disk on such a short time scale do not translate
into the mass transfer rate, which changes only on time scales of the
order the ratio of the scale height of the stellar atmosphere to its
radius times $t_{\rm w}$. The mass transfer rate is therefore kept
fixed at its current value when the instability starts
($\dot{M}_2\approx 2.4\times 10^{-7} \mpy$).

The instability would therefore only show up, if at all, as a
brightening of the system in the near-infrared.  The CB disk can
indeed become very bright with a bolometric luminosity increasing by
almost an order of magnitude to about $2 \times
10^{34}$~erg~s$^{-1}$. The bolometric luminosity is calculated
assuming the optically thick regions of the disk radiate as a
blackbody and that the disk is seen face-on. Fig.~8 shows the
evolution of the CB disk during the rise to the high luminosity
state. A front develops at the inner unstable radius and propagates
outward, transporting matter to larger radii and gradually heating the
inner regions to mid-plane temperatures reaching $10^5$~K. This is
analoguous to an outside-in outburst in a dwarf novae accretion disk,
where the outer (tidally truncated) radius becomes unstable first and
a front propagates in the direction of the flow. However, in the CB
case, the front stalls quickly due to the combination of the higher
critical density needed to heat the disk and lower available density
to do this (dilution into a larger volume of the transported mass). In
contrast, a dwarf novae outside-in front heats the entire accretion
disk because of the decreasing $\Sigma_{\rm max}$ and volume at
smaller radii.  At the luminosity maximum, only the inner 0.02~AU are
hot compared to the total (optically thick) CB disk size of 2.5~AU.

Our choice of boundary condition precludes the possibility of mass flow
to the secondary. This may not be exact during outburst when the inner
region becomes sufficiently thick ($H/R \gta 0.05$) that some matter may
flow from the CB disk back to the companion star (Artymowicz \& Lubow
1996).  Such flows are non axisymmetric and their inclusion is beyond the
scope of our simplified one dimensional radial description of the
vertically integrated structure. Note that these flows are unlikely to
change our conclusions on the possibility of having outbursts in CB
disks.

\section{Spectral Energy Distributions}

Since the temperatures, densities, and size of the CB disk vary with
time, the spectral appearance will depend upon the evolutionary stage
of the binary system.  The energy distributions for a CB disk viewed
normal to its surface for three model sequences are illustrated in
Figs. 9-11 assuming a distance of 100 pc. The distributions are shown
for the CB disk early in the evolution, prior to, and during the
accelerated mass transfer phase.  In these calculations the disk
surface is assumed to emit black body radiation at its effective
temperature. This should be a reasonable approximation since the
frequency dependence of the opacities (see Bell et al. 1997) show the
CB disk to be optically thick to radiation at wavelengths $\lta 350
\mu$m. The figures show the integrated flux over the optically thick
parts of the disk.

It is seen that the continuum peaks in the near-infrared at $\lambda
\sim 3 \mu$m for the models sequences during the entire binary
evolution.  There is a tendency for the flux to increase for the more
advanced evolutionary phases, when CB disks are larger and more
massive.  However, the flux does not increase at all wavelengths. The
flux at $\lambda \lta 3 \mu$m can decrease even though the mass input
rate into the CB disk increases (see Fig. 9).  This is due to the
increase of the binary orbital separation in the accelerated evolution
phase, causing the radiating surface area to increase and the
temperatures in the inner regions to decrease. These results taken
together indicate that the spectral energy distributions are not
sensitive to $\alpha$ and $\delta$ in the range of 0.001 - 0.005.

To determine the observability of such disks, the flux contributions
from the donor star and a steady accretion disk surrounding the white
dwarf star must be taken into account.  As an illustration we take an
evolution with $\alpha = 0.005$, $\delta = 0.005$, and $t_w = 2 \times
10^8$ yr at $t=5 \times 10^6$ yr.  At that evolutionary phase, a
secondary of mass equal to $0.36 \msun$ transfers matter to the white
dwarf of mass equal to $1.14 \msun$ at the rate of $1.3 \times
10^{18}$ g s$^{-1}$.  We assume that the outer radius of the accretion
disk is 0.67 times the Roche lobe radius of the white dwarf. The
luminosity and effective temperature of the secondary are $1.1 \times
10^{-3} \lsun$ and 3500 K respectively, the values for a main sequence
star of the same mass.  Assuming a main sequence secondary is likely
to overestimate the contribution from the donor star, which is not in
thermal equilibrium because of the mass transfer.  In Fig. 12 the
spectral energy distribution from a steady state accretion disk and
the low mass secondary component are included with the contribution
from the CB disk.  It is evident that the contribution of the CB disk
increases with wavelength, and exceeds that due to the accretion disk
and donor by a factor of about 20 and 400 at 10 $\mu$m and $350 \mu$m
respectively.  For wavelengths where the CB disk dominates the
wavelength dependence is, $F_{\lambda} \sim \lambda^{-2.6}$.  The
radius enclosing 90\% of the total flux emitted by the CB disk is
$\sim 5 \times 10^{11} - 1.6 \times 10^{12}$ cm at $10 \mu$m and
$10^{12} - 10^{13}$ cm at $100 \mu$m.  Hence, the cool outer regions
of the CB disk at $R \gta 5 \times 10^{11}$ cm provide the infrared
and submm waveband signatures of its presence.  We point out that the
neglect of irradiation effects by the secondary is not expected to
significantly modify this result since its importance in the outer
parts of the disk 
(as inferred from the variation of $H/R$ with respect to radius) 
is diminished by the shadowing effect by the
innermost regions. We note, however, that the actual height of the 
disk differs from that estimated by the pressure scale height and 
irradiation effects of the outermost regions could be important 
(see Chiang \& Goldreich 1997).  In this case the infrared and sub mm
fluxes would be increased and, hence, the fluxes estimated from the 
CB disk in the absence of irradiation effects should be considered a 
lower bound.

\section{CONCLUSIONS}

The evolution of a CB disk has been investigated for cases where it
significantly accelerates the binary evolution, such that the mass
transfer rate onto the white dwarf can reach $\mdot \gta 10^{-8}
\mpy$, representative of nova-like variables and super soft
sources. The column densities and temperatures in these disks are
found to be similar to those in circumstellar (CS) disks surrounding
young stellar objects.  The column densities decrease with radius and
reach maxima in the innermost regions of the disk with $\Sigma_{max}$
ranging up to $4 \times 10^4$~g~cm$^{-2}$.  The midplane temperatures
range from $\sim 3500$ K at the inner edge to less than about 30 K in
the optically thin outer regions.  In contrast to the CS disks, where
accretion can take place steadily, the outflowing mass in CB disks is
distinctly nonsteady.  The outer boundary of the CB disk, $R_d$,
increases with time (roughly as $R_d \sim t^{1/2}$).  We find that CB
disks around CVs should extend to several AU, about 100 times smaller
than CS disks.

The numerical results show that the fractional mass input rate into
the CB disk required to accelerate the evolution is lower than found
in Taam \& Spruit (2001). This is due to the effect of convection on
the disk structure (ignored in Taam \& Spruit), which causes the lower
stable branch of the $\nu-\Sigma$ relation to extend to much higher
column densities.

We find that the CB disks can become thermally unstable, but only
under rather special conditions, namely at high mass transfer rates
combined with short orbital periods. Such conditions could be met in
the short period double degenerate AM CVn systems or, in a different
context, in disks around Be stars (Lee et al. 1991). For normal CVs,
we find that the CB disk is unlikely to become unstable because of the
long timescale needed to reach the critical column density.  However,
depending upon the response of the secondary to mass loss, the CB disk
could become unstable during the accelerated evolution. If the CB disk
is unstable, a heat front develops and propagates outward from the
inner radius. In contrast to the accretion disks in dwarf novae, the
heating front is quenched quickly because of the difficulty in raising
the lower column densities encountered at larger radii to the
increasing critical column densities needed to sustain
propagation. The inner region cycles between the hot and cold states
on a timescale of tens of years, inducing changes in the bolometric
luminosity of a factor $\sim$ 10.

We find that the spectral energy distributions expected of CB disks in
CVs can dominate the emission from the donor star and the accretion
disk of the white dwarf at wavelengths $\lambda \gta 3 \mu$m. At
longer wavelengths the relative contribution from the CB disk to the
total emission from the system increases.

Typical flux levels for a nova-like variable for an assumed distance
of 100 pc would be about 0.3 Jy at 10 $\mu$m and a factor of 10 lower
at $450 \mu$m. Although the distances to CV's are not well known, the
model calculations suggest that even in nova-like variables with
distances order of 1 kpc, detection of CBs in the 10 - 20 $\mu$m
region should be possible using the infrared detectors on the Gemini
Observatory (Telesco et al.  2001) and the Keck Observatory (Jones et
al. 1998). Although the emission from these CB disks becomes optically
thin at 350 $\mu$m, it is possible that the longer wavelength
radiation at $450 \mu$m may be detectable using the SCUBA instrument.

In our interpretation the nova-like variables would be CVs with mass
transfer accelerated by a CB disk, which makes them natural targets
for searches for CB disks. However, the predicted emission from the CB
disk before the accelerated phase is similar, for a substantial period
of time, to that during the accelerated phase.  The possibility for
their detection is therefore not limited to the systems with high mass
transfer rates. We speculate that the `extra infrared components'
discovered by Harrison et al. (2000) in several well studied dwarf
novae may in fact be due to CB disks. These observations show that
searches of CB disks surrounding bright cataclysmic variable systems
are technically possible with current instrumentation.

\acknowledgements

\n We thank T. Currie for his assistance in this study and
J.-P. Lasota for useful discussions.  In addition, we thank the
referee, Dr. Kristen Menou, for his report which has improved the
clarity of the manuscript.  This research was supported in part by the
National Science Foundation under Grant No. AST-9727875 and by NASA
under the National Space Grant College and Fellowship Program. GD
acknowledges support from NASA grants NAG 5-7007 and NAG 5-7034. HS
acknowledges support from the European Commission under TMR grant
ERBFMRX-CT98-0195 (`Accretion onto Black Holes, Compact Objects and
Protostars').



\begin{figure}
\epsfig{file=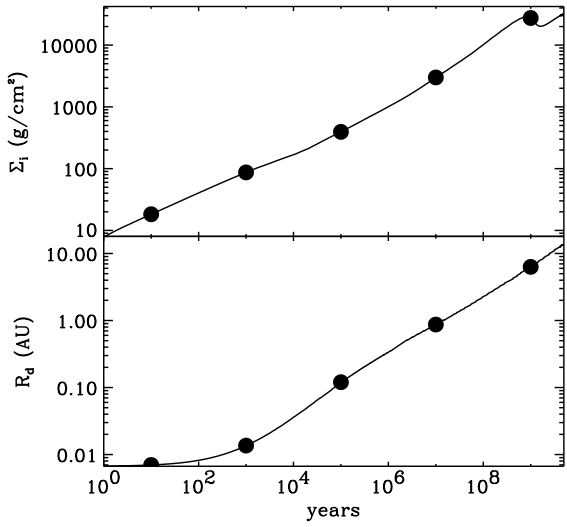,width=3in}
\caption{The evolution of a model CB disk with constant mass input
rate (
$\dot{M}_{\rm i}=-10^{14}$~g~s$^{-1}$) and inner radius ($R_{\rm
i}=10^{11}$~cm), for viscosity parameter $\alpha= 0.001$. Top panel:
column density at the inner disk edge $\Sigma_{\rm i}$. Bottom panel:
outer radius of the disk defined as the radius where the optical depth
$\tau=1$. The radial structure at the times indicated by the dots on
these curves are presented in Fig.~2.  The surface density required
for disk-instability ($\Sigma_{\rm max}\approx 6\times
10^4$~g~cm$^{-2}$ in this case) is not reached.}
\end{figure}

\begin{figure}
\epsfig{file=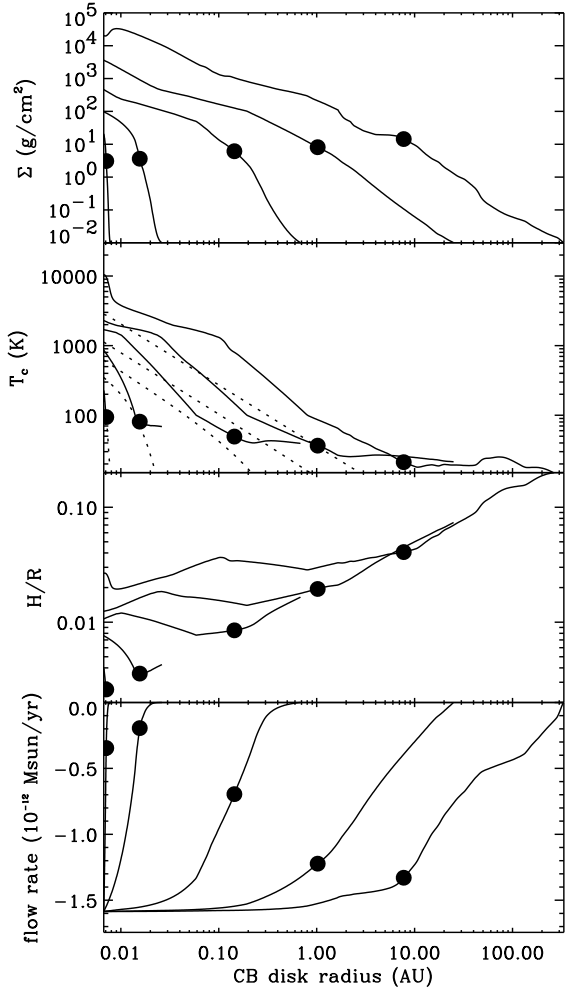,width=3in}
\caption{The radial structure of the CB disk presented in Fig.~1 at
different times during its evolution. The column density, $\Sigma$,
mid-plane temperature $T_{\rm c}$ (solid line), effective temperature
$T_{\rm eff}$ (dashed line), ratio of pressure scale height to radius,
$H/R$, and mass flow rate $\dot M$ (in units of $10^{-12} \mpy$,
negative
values indicate mass is flowing away from the binary)
are
illustrated as a function of radius and time (each radial curve
corresponds to a dot in Fig.~1).  The radius at which the optical
depth equals 1 is marked by a dot on each curve.}
\end{figure}

\begin{figure}
\epsfig{file=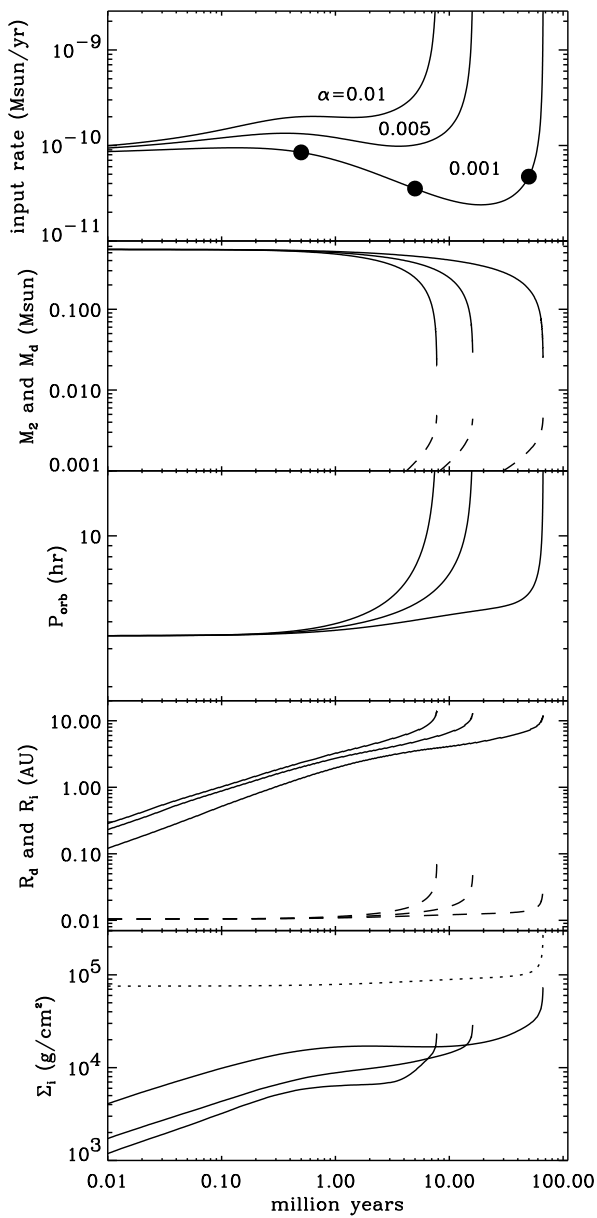,width=3in}
\caption{Evolution of the CB disk with the evolution of and angular
momentum loss from the binary system taken into account. Three
calculations are shown in each panel for values of the viscosity
parameter $\alpha=0.01$, 0.005 and 0.001. The calculations start with
a secondary mass of 0.55 $\msun$ and a white dwarf of 0.95 $\msun$.
Viscosity parameter is $\alpha=0.001$, time scale for angular momentum
loss by magnetic braking of the binary is $t_{\rm w}=2\times
10^8$~years, and mass fed into the CB disk $-\dot{M}_{\rm i}$ is a
fraction $\delta=0.005$ of the mass transfer rate.  From top to
bottom: the mass input rate into the CB disk $-\dot{M}_{\rm i}$; the
mass of the secondary and the mass of the disk (dashed line); the
orbital period; the outer radius of the CB disk (at which the optical
depth equals 1) and the inner radius of the CB disk (dashed line); the
column density at the inner disk edge $\Sigma_{\rm i}$ and the
corresponding critical column density $\Sigma_{\rm max}$ above which
local thermal instabilities are expected (dashed line). The radial
structure of the CB disk is shown at different times (indicated by the
dots in the first panel) in Fig.~4.}
\end{figure}

\begin{figure}
\epsfig{file=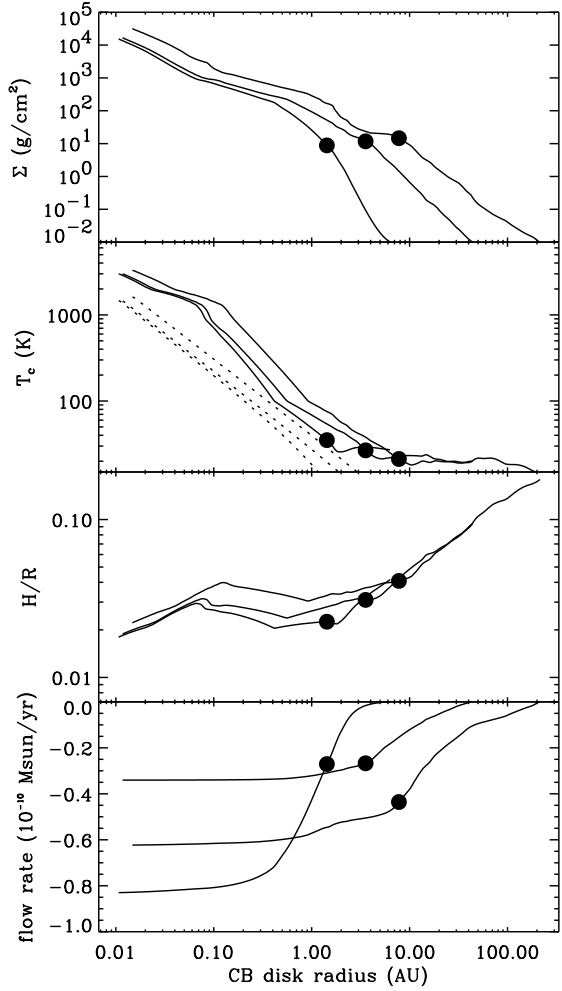,width=3in}
\caption{The radial structure of the CB disk with $\alpha=0.001$
presented in Fig.~3, at different times during its evolution.  The
column density $\Sigma$, the mid-plane temperature $T_c$ (solid line)
and the effective temperature $T_{\rm eff}$ (dashed line), the ratio
of the pressure scale height to radius $H/R$, and the mass flow rate
$\dot M$ (in units of $10^{-10} \mpy$) are shown as functions of
radius and time (at $t=5\times [10^5, 10^6, 10^7]$ years, see Fig.~3).
The radius at which the optical depth $\tau=1$ is shown as a dot on
each curve.}
\end{figure}

\begin{figure}
\epsfig{file=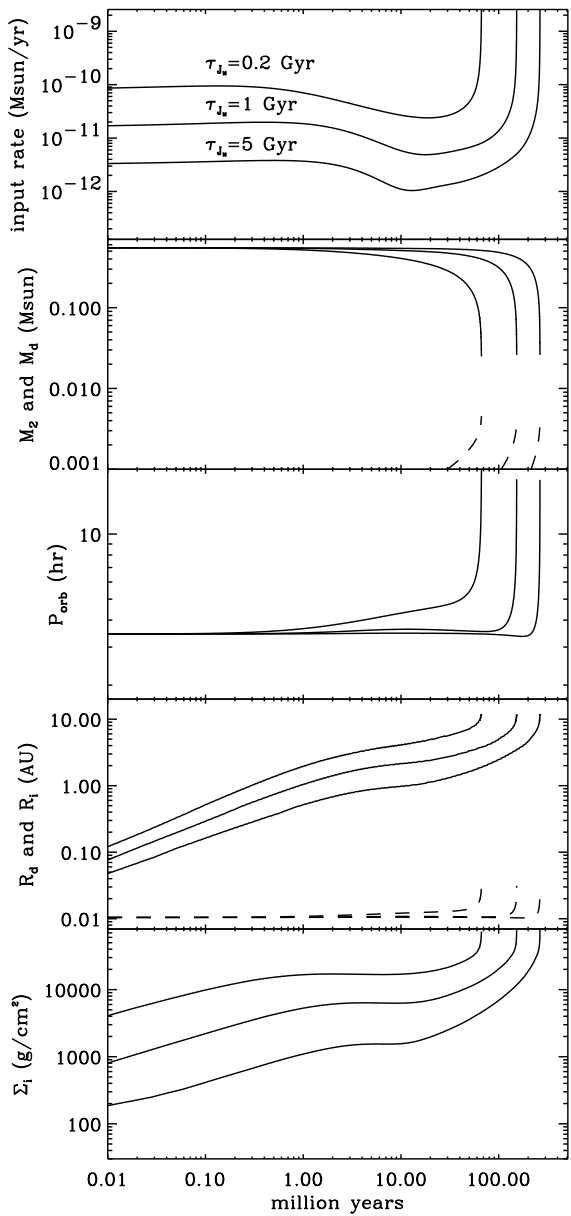,width=3in}
\caption{Same as Fig.~3 but showing the influence of varying $t_{\rm
w}$, the timescale for binary angular momentum loss by magnetic
braking.
The other parameters are $\alpha=0.001$ and $\delta=0.005$
and the calculations start with a secondary mass of 0.55 $\msun$ and a
white dwarf of 0.95 $\msun$.}
\end{figure}

\begin{figure}
\epsfig{file=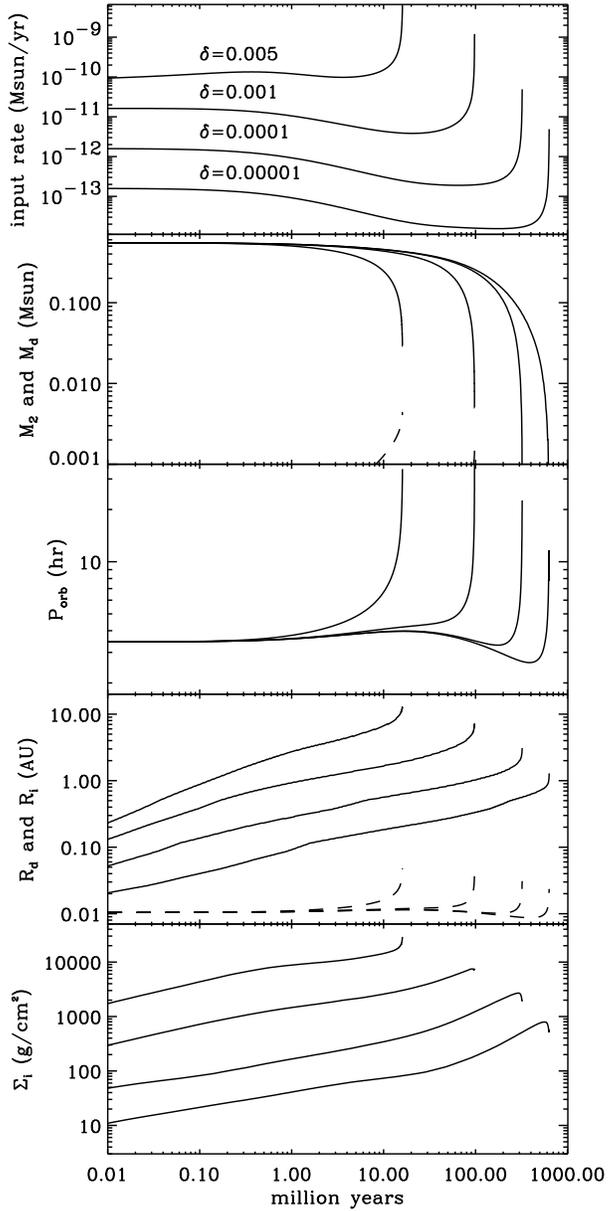,width=3in}
\caption{Same as Fig.~3 but showing the influence of varying $\delta$,
the fraction of the binary mass transfer rate which is input in
the CB disk.  The other parameters are $\alpha=0.005$ and $t_{\rm
w}=2\times 10^8$~years and the calculations start with a secondary
mass of 0.55 $\msun$ and a white dwarf of 0.95 $\msun$.}
\end{figure}

\begin{figure}
\epsfig{file=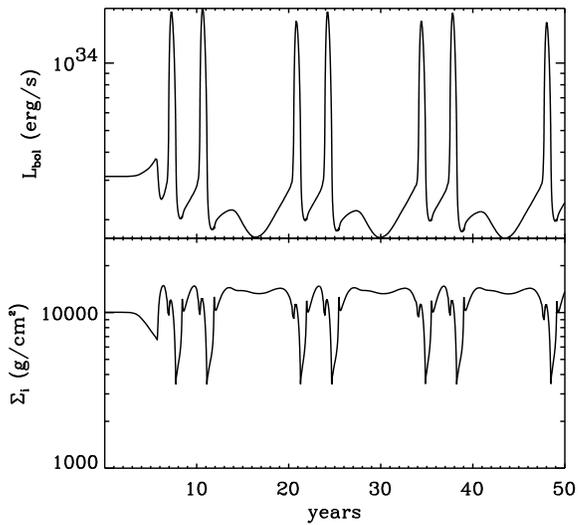,width=3in}
\caption{Example of a thermally and viscously unstable CB disk. The
calculation starts with a $0.55 \msun$ secondary and a $0.95 \msun$
primary. We assume the secondary {\em contracts} in response to mass
transfer (see \S4.3), $\alpha_{\rm cold}=0.005$, $\alpha_{\rm
hot}=0.01$ and $\delta=0.005$. The evolution up to $t\approx 30$~Myr
is comparable to the model shown in Fig.~3. During the runaway, the
radius of the secondary shrinks to 0.16 $\rsun$, decreasing the
orbital period to $\approx 2$ hours and bringing the CB disk to
shorter radii (0.007 AU). The critical $\Sigma_{\rm max}$ is then low
enough for the disk to become unstable.  The upper and lower panels
show the bolometric luminosity and the column density at the inner
edge of the CB disk after it becomes unstable. The mass input into the
disk is constant during the cycles at $-\dot{M}_{\rm i}\approx
10^{-9}\mpy$. The evolution of the radial structure of the CB disk for
the outburst at $t\approx 24$ yrs is shown in Fig.~8.}
\end{figure}

\begin{figure}
\epsfig{file=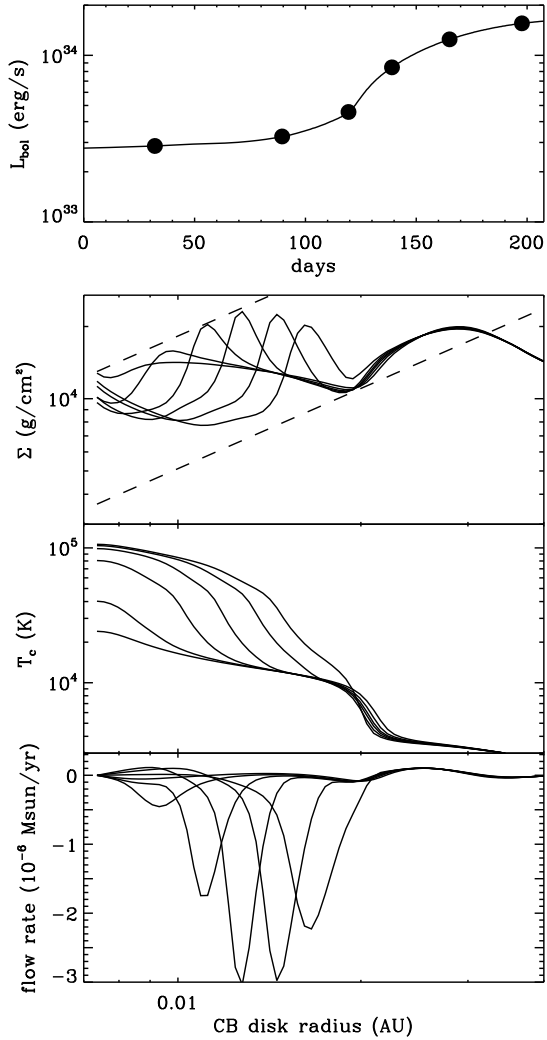,width=3in}
\caption{Evolution of the radial structure of a CB disk during the
rise to a luminous state. The model shown is that of Fig.~7 for the
outburst peaking at $t\approx 24$~yrs.  The top panel is a zoom on the
bolometric luminosity lightcurve with the dots indicating at which
times the 6 radial structures shown underneath were taken. The bottom
three panels show the radial profiles of the column density $\Sigma$,
mid-plane temperature $T_{\rm c}$ and mass flow rate in the CB disk
$\dot{M}$ (in units of $10^{-6} \mpy$).  The two dashed lines in
column density panel are the critical densities $\Sigma_{\rm max}$
(top) and $\Sigma_{\rm min}$ (bottom). In contrast to the outbursts in
standard {\em accretion} disks, the front does not propagate very far
out into the CB disk before it stalls.}
\end{figure}

\begin{figure}
\epsfig{file=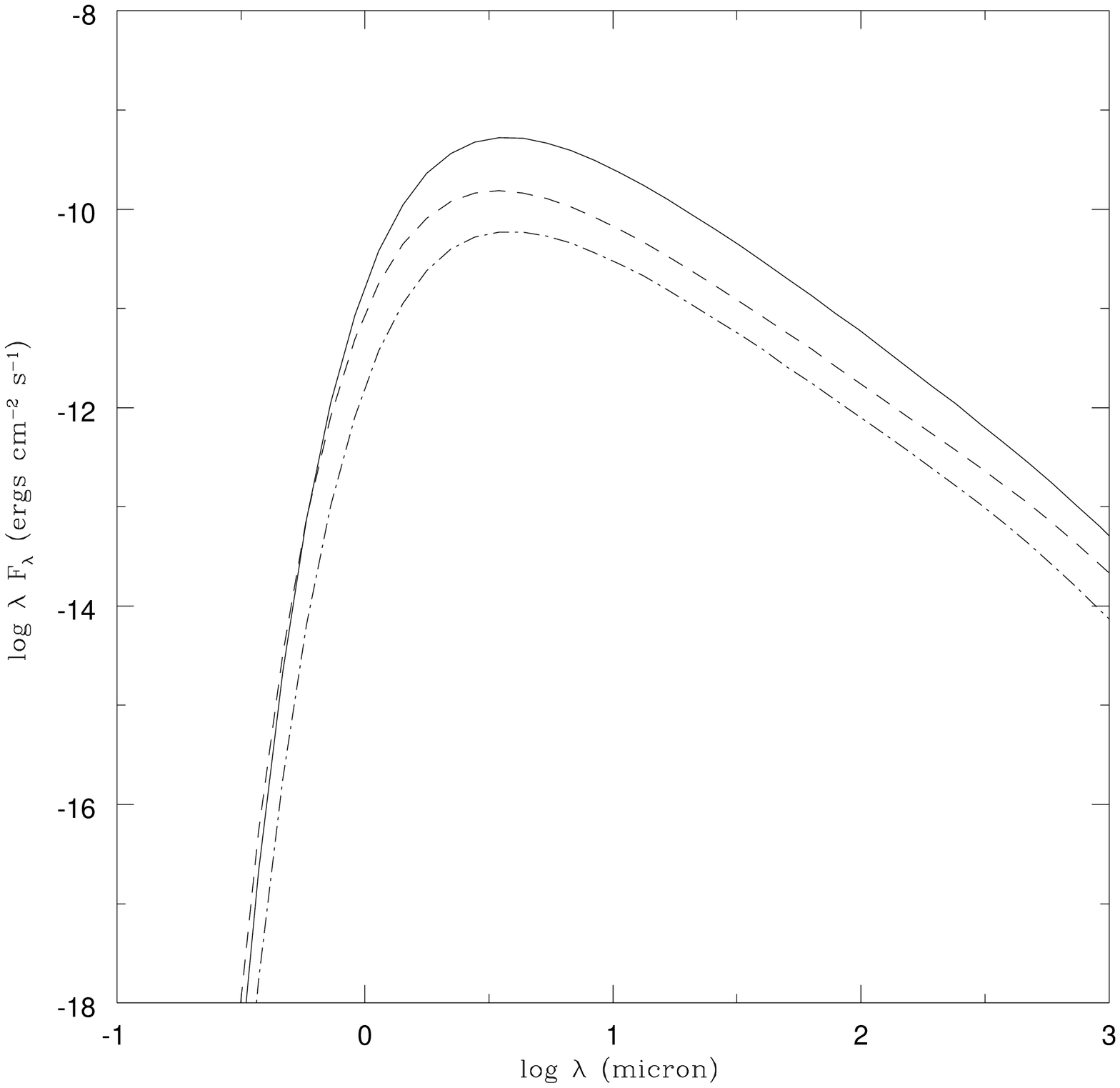,width=3in}
\caption{The spectral energy distribution for a CB disk
for the model sequence with $\alpha = 0.001$ and
$\delta = 0.005$ for an assumed distance of 100 pc.
The dash - dot curve, dashed curve, and solid curve show
stages of evolution at $t=5 \times 10^6$ yr,
$5 \times 10^7$ yr, and $7 \times 10^7$ yr respectively.}
\end{figure}

\begin{figure}
\epsfig{file=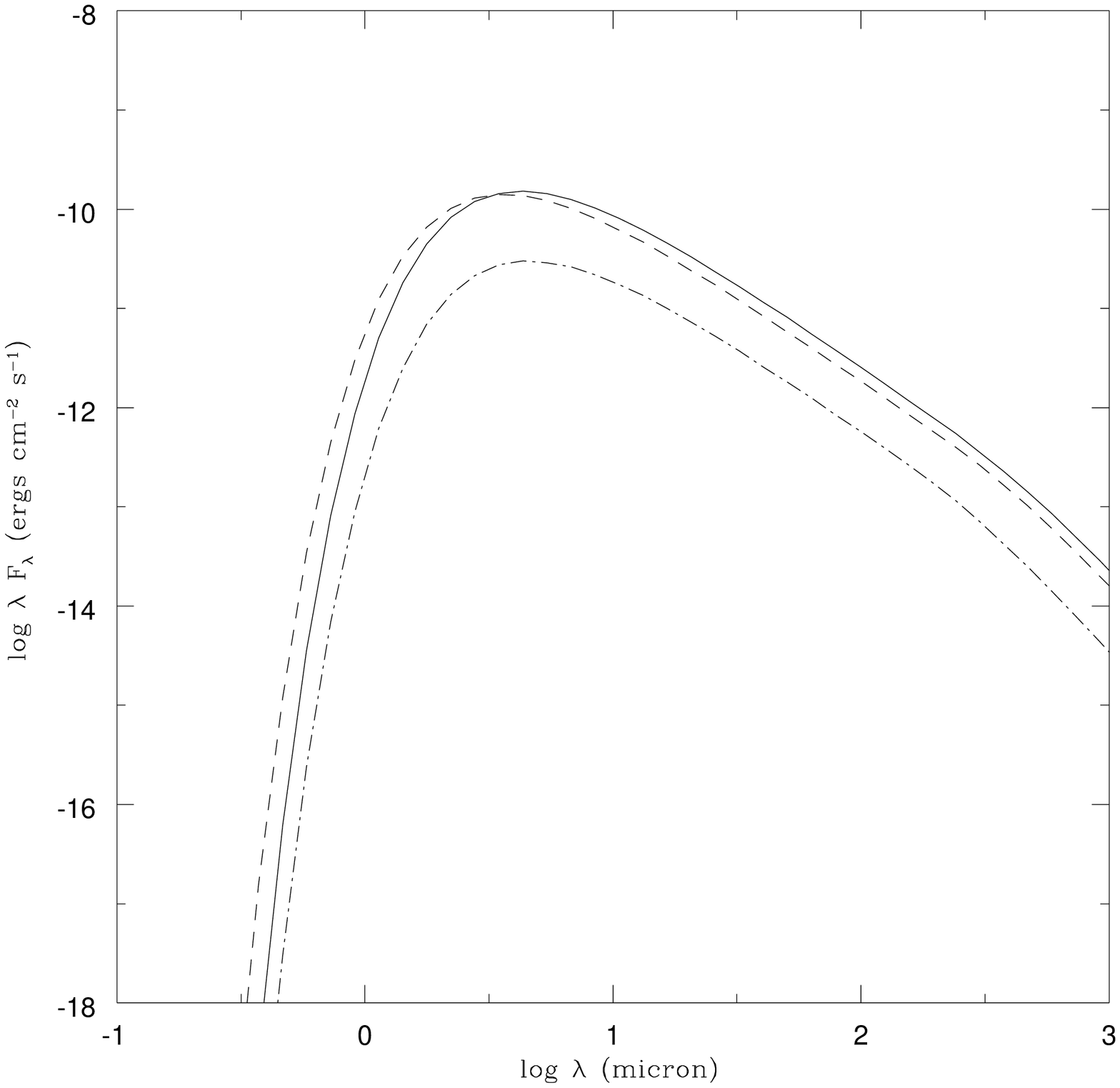,width=3in}
\caption{The spectral energy distribution for a CB disk
for the model sequence with $\alpha = 0.005$ and
$\delta = 0.001$ for an assumed distance of 100 pc.
The dash - dot curve, dashed curve, and solid curve show
stages of evolution at $t=8 \times 10^6$ yr,
$8 \times 10^7$ yr, and $9 \times 10^7$ yr respectively.}
\end{figure}

\begin{figure}
\epsfig{file=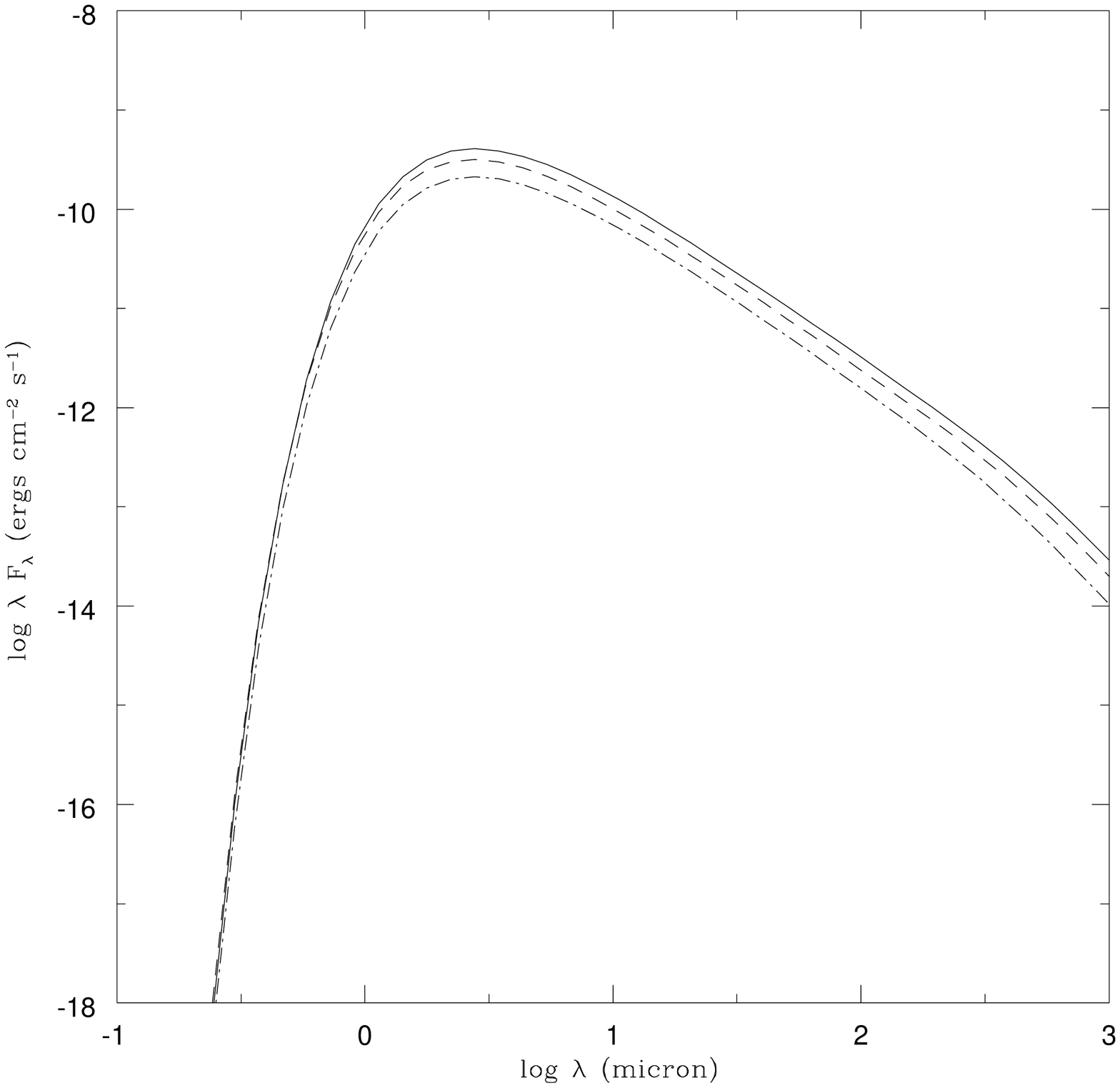,width=3in}
\caption{The spectral energy distribution for a CB disk
for the model sequence with $\alpha = 0.005$ and
$\delta = 0.005$ for an assumed distance of 100 pc.
The dash - dot curve, dashed curve, and solid curve show
stages of evolution at $10^6$ yr,
$5 \times 10^6$ yr, and $9 \times 10^6$ yr respectively.}
\end{figure}

\begin{figure}
\epsfig{file=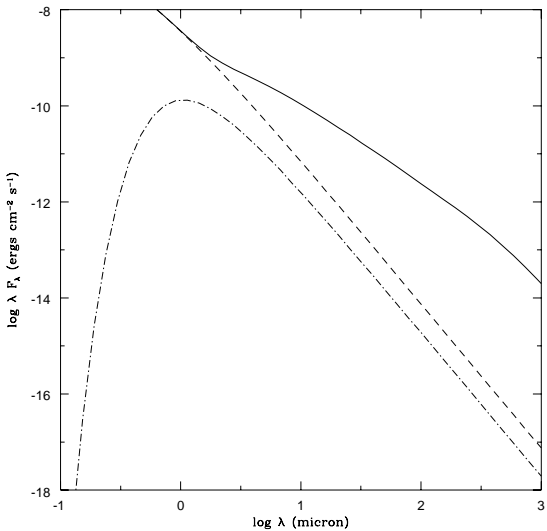,width=3in}
\caption{The spectral energy distribution for a cataclysmic
variable system with CB disk, for $\alpha = 0.005$ and $\delta =
0.005$ at an evolutionary time of $5 \times 10^6$ yr. At this time,
$\dot M_2$, is $1.3 \times 10^{18}$ g s$^{-1}$.  The distance is
assumed equal to 100 pc.  The dash - dot curve shows the contribution
from a low mass secondary star of mass $0.36 \msun$, the dashed curve
corresponds to the sum of the contribution from a steady state
accretion disk surrounding the white dwarf component of mass $1.14
\msun$ and the low mass secondary, and the solid curve includes the
contribution from the CB disk as well.  Note that the CB disk
dominates the spectral energy distribution for $\lambda \gta 3 \mu$m.}
\end{figure}
\end{document}